\documentclass[]{aastex701}

\def\yr4{2024 YR$_4$}

\def\las{\mathrel{\hbox{\rlap{\hbox{\lower3pt\hbox{$\sim$}}}\hbox{\raise2pt\hbox{$<$}}}}}
\def\gas{\mathrel{\hbox{\rlap{\hbox{\lower3pt\hbox{\sim$}}}\hbox{\raise2pt\hbox{$>$}}}}}
\received{June 12, 2025}
\revised{August 8, 2025}
\accepted{August 11, 2025}
\submitjournal{Astrophysical Journal Letters}

\begin{document}

\title{The Potential Danger to Satellites due to Ejecta from a 2032 Lunar Impact by Asteroid \yr4}

\author[0000-0002-1914-5352]{Paul Wiegert}
\affiliation{Department of Physics and Astronomy \\
The University of Western Ontario \\
London, Canada}
\affiliation{Institute for Earth and Space Exploration (IESX) \\
The University of Western Ontario \\
London, Canada}
\email{pwiegert@uwo.ca}
\author[0000-0001-6130-7039]{Peter Brown}
\affiliation{Department of Physics and Astronomy \\
The University of Western Ontario \\
London, Canada}
\affiliation{Institute for Earth and Space Exploration (IESX) \\
The University of Western Ontario \\
London, Canada}
\email{pbrown@uwo.ca}
\author[0009-0000-7999-3241]{Jack Lopes}
\affiliation{Department of Physics and Astronomy \\
The University of Western Ontario \\
London, Canada}
\affiliation{Institute for Earth and Space Exploration (IESX) \\
The University of Western Ontario \\
London, Canada}
\email{jlopes22@uwo.ca}
\author[0000-0003-0634-9599]{Martin Connors}
\affiliation{Centre for Science \\
Athabasca University \\
Athabasca, Alberta, Canada}
\affiliation{Department of Physics and Astronomy \\
The University of Western Ontario \\
London, Canada}
\affiliation{Institute for Earth and Space Exploration (IESX) \\
The University of Western Ontario \\
London, Canada}
\email{martinc@athabascau.ca}

\correspondingauthor{Paul Wiegert}

\begin{abstract}
On 2032 December 22 the 60m diameter asteroid \yr4 has a 4\% chance of impacting the Moon. Such an impact would release 6.5 MT TNT equivalent energy and produce a $\sim$ 1 km diameter crater. We estimate that up to 10$^{8}$ kg of lunar material could be liberated in such an impact by exceeding lunar escape speed. The current overall probability is about 1\% that the asteroid will impact the Moon at a location such that more than 10\% of the ejected material would accrete to the Earth on timescales of a few days. If this were to occur, the lunar ejecta-associated particle fluence at 0.1 - 10 mm sizes could produce upwards of years to of order a decade of equivalent background meteoroid impact exposure to satellites in near-Earth space late in 2032. Our results demonstrate that planetary defense considerations should be more broadly extended to cis-lunar space and not confined solely to near-Earth space.
\end{abstract}



\section{Introduction} \label{sec:intro}

Asteroid \yr4 is a 60 m diameter asteroid \citep{Rivkin_2025_JWST_2025} that is confidently expected to miss the Earth but that as of this writing has a 4\% probability of striking the Moon at 13 km/s on 2032 December 22. Here we report on the amount of lunar material potentially injected into cis-lunar space should that impact occur, and the implications for Earth's satellite constellations in particular. Of primary concern are ejecta particles above the impact hazard threshold (0.1~mm) for satellites. This is the size range identified by NASA \citep{Cooke_2017} as being at the limit of causing substantial damage to satellites, with particles smaller than this size normally leading to surface degradation alone.

Our goal in this work is to provide an order of magnitude estimate of the expected short-term effect of lunar ejecta on near-Earth space from a potential impact of \yr4 in late December 2032. We emphasize that there exist order of magnitude uncertainties in the following analysis. This is particularly true in regard to ejecta size-frequency distributions at small sizes and the mass fraction of ejecta able to exceed lunar escape speed.  

To assess the particle population which might reach Earth should \yr4 impact the Moon we need to estimate:
\begin{enumerate}
    \item The size of the crater produced in the impact
    \item The amount of material ejected in the impact which has speeds above lunar escape
    \item The size frequency distribution of the escaping ejecta
    \item The range of locations on the moon where an impact might occur
    \item The delivery efficiency of escaping ejecta to near-Earth space.
\end{enumerate}

This letter is structured as follows:
In section \ref{sec:crater} we address items 1-3. Items 4-5 are addressed in Sections \ref{sec:methods} and \ref{sec:efficiency}. Finally, we estimate the fluence of lunar particles with D $>$ 0.1 mm from a \yr4 impact and compare to the background meteoroid flux/hazard in section \ref{sec:effects}.

\section{Crater production and Ejecta Properties} \label{sec:crater}

In what follows we adopt the scaling relations for transient crater size, ejected mass and velocity distribution summarized and described in \cite{Holsapple_Schmidt_1982,Housen_Schmidt_Holsapple_1983, Holsapple_Schmidt_1987, Holsapple_1993, Housen_Holsapple_2011}.

Adopting the JWST determined diameter of 60 m for \yr4 \citep{Rivkin_2025_JWST_2025}, assuming a bulk density for \yr4 of 3000 kg~m$^{-3}$ and an impact speed of 13 km/s (appropriate for all locations along the potential lunar impact corridor; see section \ref{sec:methods}) we may estimate a crater size. Using $\Pi$-Scaling summarized in \cite{Holsapple_1993} appropriate for a target of hard rock in the gravity-dominated regime we adopt  cratering scaling exponents of $\mu_{crater}$=0.22 and $k_{crater}$=1.0 for a zeroth order crater diameter estimate \citep{Housen_Holsapple_2011}. We choose hard rock as the fine regolith is expected to be no more than 10~m in depth \citep{watkins_boulder_2019}. 

Under these assumptions, the impact will produce a transient crater of 1.4~km diameter. If instead we assume the impact is more appropriate to loose soil/fine regolith we find a transient crater diameter of 0.7 km (\cite{Holsapple_Schmidt_1982}, Table 1, case 5). Given the various uncertainties, we will assume that the impact of \yr4 into the moon will produce a transient crater of order 1 km in diameter. For context, based on the lunar crater production flux of \cite{Neukum_Ivanov_Hartmann_2001} a 1 km diameter crater forms every $\sim$ 5000 years. This emphasizes how unusual/rare a potential lunar impact is for an object as large as \yr4, at least on human time scales if not on geological ones.

In general, impacts on the Moon produce ejecta, a portion of which may escape lunar gravity. \citet{Melosh_1985} showed that near surface spallation during lunar impacts can loft a small portion ($\sim$0.01\%) of material above escape speeds. Subsequent work \citep{Housen_Holsapple_2011} showed that most of the mass of high velocity ejecta is in the form of $\micron$ to mm-sized fragments.  

For a 1 km diameter crater, we expect $\sim$ 10$^{11}$ kg of mass to be displaced from the crater. Using the ejecta velocity formalism of \citet{Housen_Holsapple_2011} and an ejecta -velocity exponent $\mu$=0.41 appropriate for sandy material we find that only 0.02\% - 0.2\% of ejecta exceed lunar escape speed, implying release of $\sim$ 10$^{7-8}$ kg during the impact.

The size frequency distribution (SFD) of fines produced during lunar impacts is poorly constrained, with most work focusing on the large-end tail of the ejecta distribution  \citep[e.g.][]{Singer_Jolliff_McKinnon_2020}. For simplicity we follow the methodology of \citet{Jedicke_Alessi_Wiedner_Ghosal_Bierhaus_Granvik_2025} and assume that the largest ejecta fragments will be one meter in diameter or smaller. We also adopt their assumption that a simple power-law expresses the number of fragments between the smallest fines (assumed to be a micron in diameter) and this upper size limit. Adopting this SFD of the form 

\begin{equation}
    N(>D) = CD^{-u},\label{eqn:sfd}
\end{equation}

where $N(>D)$ is the cumulative number of fragments with diameter larger than $D$, $C$ is a normalizing constant and $u$ is the cumulative SFD power-law index. Typical values for $u$ for large crater ejecta SFD are of the order 3 to 4 \citep{Bart_Melosh_2010b} a range we adopt. For an escaping ejecta mass of 10$^{7-8}$ kg, we find N($>$ 10 mm) $\sim$ 10$^{5-9}$, N($>$ 1 mm) $\sim$ 10$^{11-13}$, N($>$ 100 $\mu$m) $\sim$ 10$^{14-16}$.

\section{Impact probability} \label{sec:methods}

\begin{figure}
\plotone{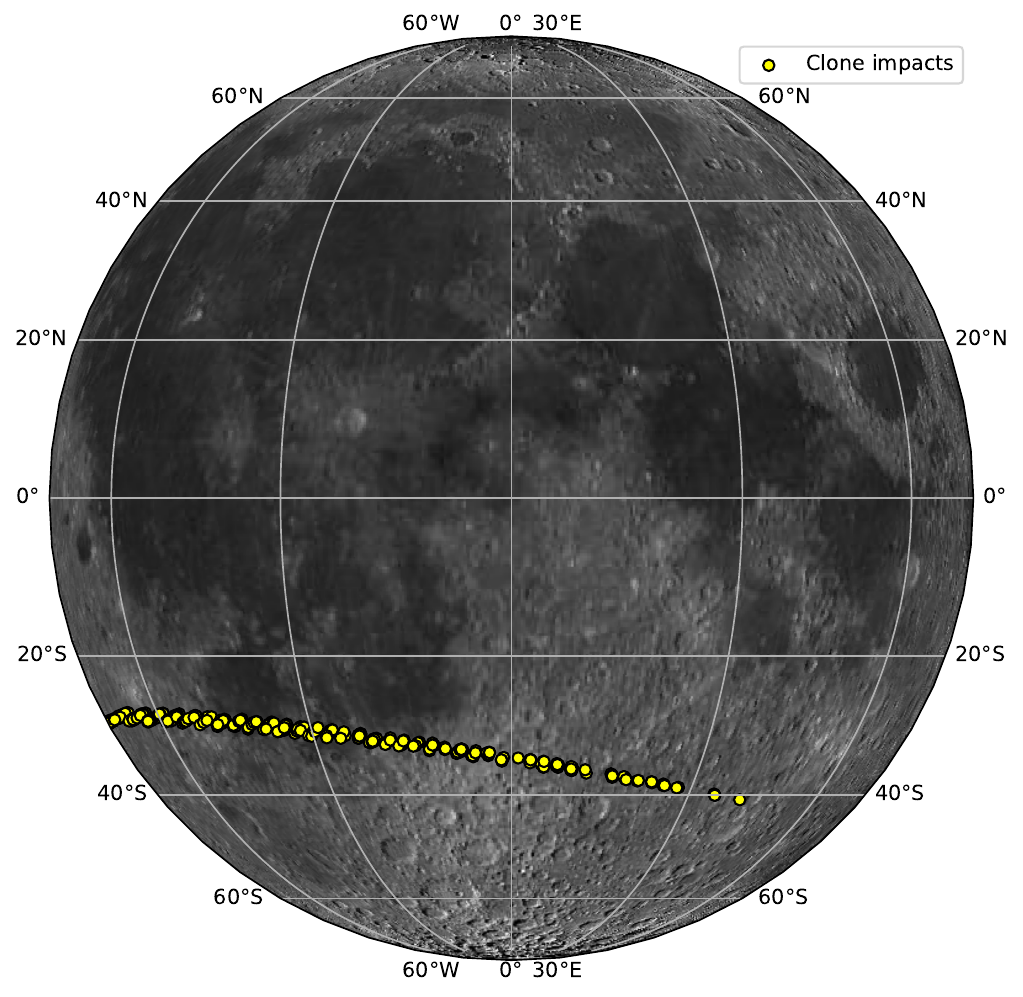}
\caption{The current impact corridor for \yr4 (yellow) projected on a map of the Moon's near side from Lunar Reconnaissance Orbiter \citep{sperobden11}. \label{fig:impactcorridor}}
\end{figure}

To determine the lunar impact probability of \yr4, asteroid trajectories were calculated via the RADAU \citep{eve85} 15th order method using an external time step of 14.4 minutes (0.01 days) up to 2032 Dec 22 1200UT (a few hours before the closest approach to the Moon) and a time step of 5 seconds from there through the encounter. The longer time step provides adequate coverage of the approach, sampling the asteroid's orbit thousands of times per heliocentric period. The shorter time step  during the encounter ---which occurs at a relative speed of 11 km/s--- allows the resolution of impact locations to within roughly 50 km on the Moon's 3476 km diameter disk. The simulations included the effects of the Sun, Moon and all the planets with their initial positions derived from the DE440 ephemeris \citep{parfolwil21}.  Radiation forces were ignored as their effects are small on particles of these sizes on these time scales. The orbital solution for \yr4 was obtained from the Center for Near-Earth Object Studies (CNEOS) Small-Body Database (SBDB) API \footnote{\url{https://ssd-api.jpl.nasa.gov/sbdb.api}} on 2025 June 5. This solution incorporates James Webb Space Telescope (JWST) observations taken on 2025 May 11, probably the last observations that can be taken before 2028, and which increased the chance of \yr4 striking the Moon slightly from 3.8\% to 4.3\%\footnote{\url{https://science.nasa.gov/blogs/planetary-defense/2025/06/05/nasas-webb-observations-update-asteroid-2024-yr4s-lunar-impact-odds/}}. Ten thousand clones were created from the covariance matrix and numerically integrated forward to their impacts, which occur within a few minutes of 2032 December 22 1520 UT, depending on the exact impacting clone.
Our simulations yield an impact probability of $4.1\pm 0.2\%$, consistent with the above-mentioned value. The impact corridor as seen from the Earth is presented in Figure~\ref{fig:impactcorridor}, that is, the 410 clones that strike the Moon are plotted. The plane of the orbit of \yr4 is well known so there is little spread across the track; the length of the track results from uncertainty in its precise location along its orbit. If an impact occurs, it will be in the southern hemisphere roughly between latitudes 30S and 40S, and largely on the Moon's leading side but with a 14\% chance of impacting on the trailing side. 

\section{Ejecta Delivery Efficiency for \yr4 impact} \label{sec:efficiency}

If an asteroid impact produced ejecta traveling in directions isotropic with respect to the Moon, then we might expect the Earth to intercept a fraction of that ejecta comparable to the fraction of the all-sky solid angle it subtends as seen from the Moon, or about $10^{-4}$. However, material ejected from a single impact on the Moon will not be ejected isotropically. The fraction that will reach the Earth directly is highly sensitive to the location of the impact. 

This sensitivity can be understood qualitatively. Because the Moon is orbiting the Earth at approximately 1 km/s, for ejected material to reach Earth quickly the impacting object needs to hit the trailing edge of the Moon in such a way that the ejected material's velocity after rising out of the Moon's gravitational well largely cancels out the Moon's orbital velocity. This leaves an ejected particle almost stationary with respect to Earth, allowing it to fall straight down toward our planet. Particles may also be ejected on hyperbolic orbits that head directly for Earth. Many of these details were worked out in the context of the delivery of lunar meteorites to Earth by \cite{glaburdun95}, though these authors were little concerned with the direct delivery of material to Earth, as lunar meteorites more often follow longer, more dynamically complex paths. \cite{kreasp14} also considered the direct delivery of material to the Earth on short time scales, in the context of searching for signs of large lunar impacts in the geological record. Other studies of lunar ejecta have primarily focused on the larger (meter) sizes and on the longer term evolution of it into mini-moons \citep{jedalewie25J}, near-Earth objects and/or Earth co-orbital asteroids \citep{jiachehua24,casmalros25}.

We will show that there is a significant probability for \yr4 to hit on a portion of the Moon with a delivery efficiency to Earth of greater than 10\% of the total material ejected, and that this debris moves quickly to near-Earth space.

\subsection{Generic delivery efficiency} \label{sec:genericeff}

\begin{figure}
\plotone{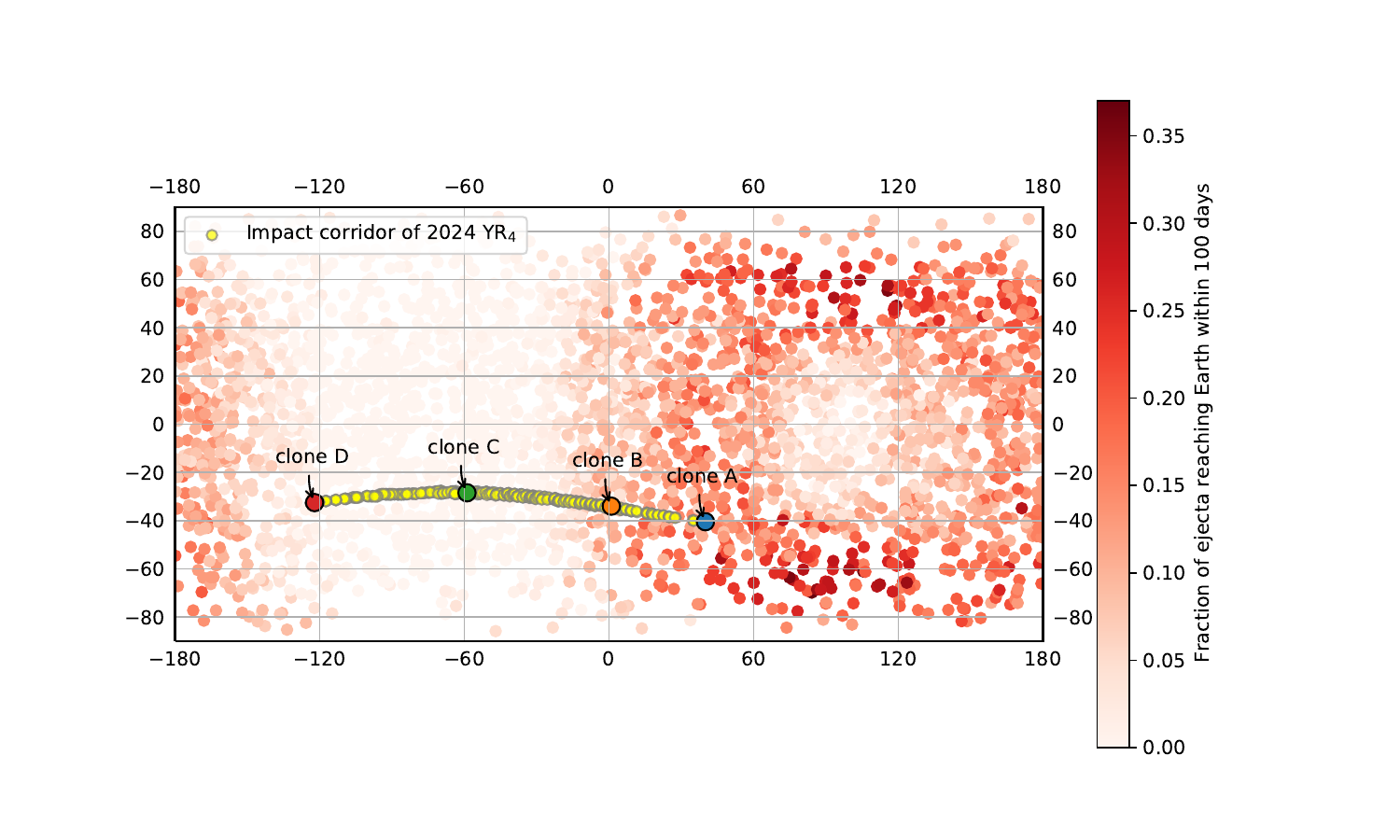}
\caption{A map of the Moon showing randomly selected impact locations colored by the fraction of escaping ejecta delivered to Earth within 100 days. The current impact corridor for \yr4 is shown in yellow with the specific impact locations examined more closely in Section~\ref{sec:expectedeff} labelled as clones A-D. See the main text for more details.\label{fig:ejecta}}
\end{figure}

Before addressing the delivery efficiencies of possible \yr4 impacts specifically, we first examine simulated hypothetical impacts at randomly selected locations across the Moon on 2032 December 22, with points colored by the fraction of material delivered to Earth within 100 days (Figure~\ref{fig:ejecta}). The figure was created from 300,000 simulated particles (3000 locations, each producing 100 ejected particles, simulation time step 0.01 days, more details below). 

Ejecta trajectories are computed with the same code as was used for computing the trajectory of \yr4 (Section~\ref{sec:methods}).  The pole and orientation of the Moon at the time of impact were computed using the SpiceyPy Python wrapper \citep{Annex2020} for NASA's SPICE toolkit \citep{Acton2018}. Specifically, we retrieved the orientation of the Moon's north pole, the libration angle of the Moon (i.e., the angle between the Earth vector and the Moon’s prime meridian), and the surface rotation speeds for Julian Ephemeris Date (JDE) 2463589.139 corresponding to approximately 1520 UT on 2032 December 22. The Moon’s North Pole direction in the ecliptic J2000 frame is given by the unit vector $\vec{n} = [0.0127,\ -0.0245,\ 0.9996]$ indicating an obliquity relative to the ecliptic of $1.6^\circ$. The libration angle was found to be $0.696^\circ$ East, meaning the sub-Earth point was slightly east of the Moon’s prime meridian at the time. The rotational velocity of the lunar surface is found to be only a few meters per second, negligible compared to the ejection speeds being considered and the rotational speed of the Moon's surface was ignored in the ejecta simulations.

The delivery efficiency to the Earth is sensitive to the precise direction and speed at which the material leaves the lunar surface. Here we adopt a single speed for ejected material, chosen just above the escape speed because this is what is expected of the bulk of the escaping material \citep{vic87}. We also adopt ejection on a cone at $45^o$ to the local normal \citep{mel89}. Specifically, each impact is taken to eject material at 2.6 km/s (just above the Moon's escape speed of 2.38 km/s) on a cone with an opening angle ranging from $45\pm5^o$ to the local normal of the geoid \citep{mel89}. 

Ejecta particles are numerically integrated with a time step of 0.01 days (= 14.4 minutes) until they hit the Moon, Earth, depart the Earth's Hill sphere (that is, reach a distance larger than 0.01~au from Earth) or the 100 day time limit expires.  From Figure~\ref{fig:ejecta} it can be seen that if the impact is on the leading side, very little or no material reaches Earth during the time frame in question; however, the delivery efficiencies can reach over 30\% on the trailing side.  

The delivery efficiency to Earth does depend on the ejection speed, and our choice of 2.6 km/s is near the optimum value for randomly distributed impacts. In Figure~\ref{fig:ejecta}, (which assumes a 2.6 km/s ejection speed) 9.7\% of impacts have delivery efficiencies greater than 10\% to Earth. This drops to only 5\% of impacts at ejection speeds of 2.55 or 2.65 km/s, and to less than 1\% at 2.52 km/s and 2.81 km/s. However, with changing ejection speed there is also a change in the pattern of high-efficiency impact locations. The pattern narrows and shifts away from \yr4's impact corridor at lower ejection speeds (decreasing overall delivery efficiency), while it broadens and shifts towards the impact corridor at higher ones (which increases efficiency). Our choice of 2.6 km/s is likely optimistic but although a more sophisticated approach could be taken, it is probably not justified until (and if) an impact of \yr4 on the Moon is better constrained.

The highest delivery rates would occur if \yr4 hit the trailing side of the Moon, though delivery efficiencies even along the midline of the Moon can be substantial. The impact corridor is overplotted on Figure~\ref{fig:ejecta}: 81 of 410 clones that impact the Moon strike in regions with greater than 10\% delivery efficiency to the Earth. Thus if \yr4 were to strike the Moon, then there is a 20\% chance that the delivery efficiency will be 10\% or higher.  Together there is a joint 0.8\% chance of \yr4 1) striking the Moon and 2) impacting in a region of high delivery efficiency. The odds of such a scenario well exceed the usual threshold of concern for asteroid Earth impact warning systems which is one in a million \citep{milchesan05,roafarche21} suggesting that though the risk is low, it is not completely negligible.

\subsection{Expected delivery efficiency from a \yr4 lunar impact} \label{sec:expectedeff}

\begin{figure}
\plotone{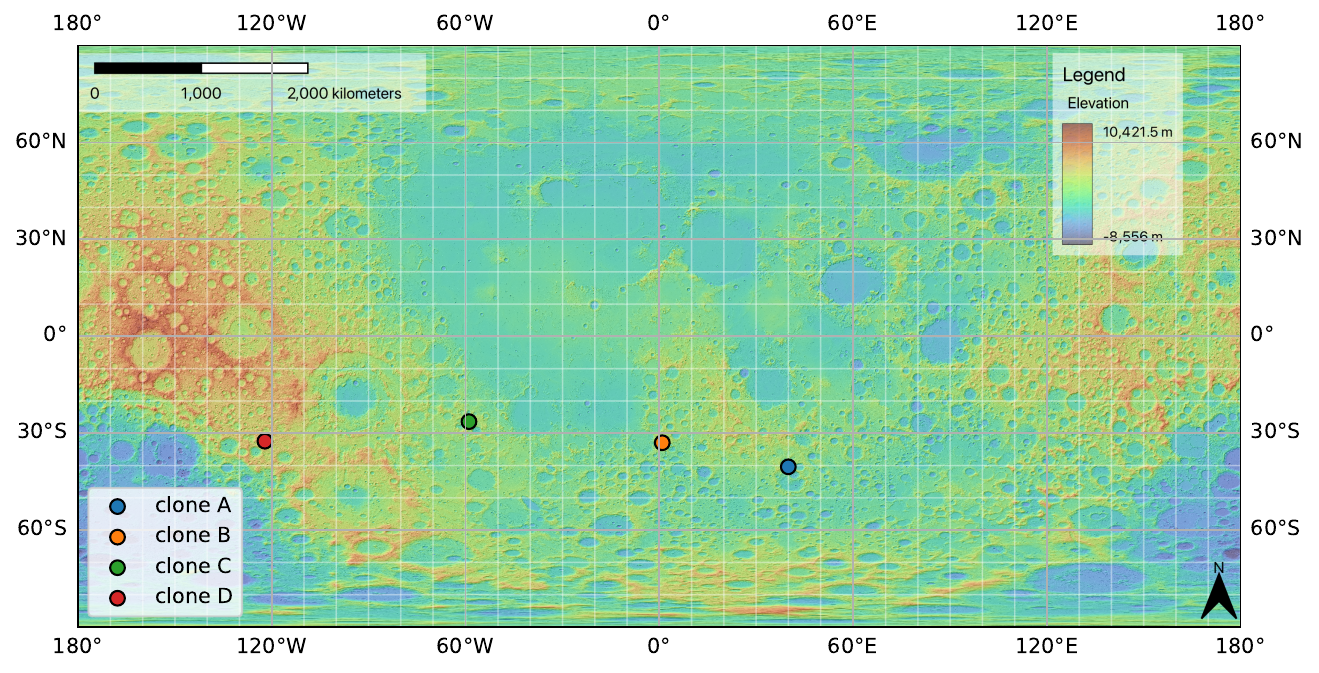}
\caption{An elevation map of the Moon showing the locations of clones A through D selected for a more detailed study of their delivery efficiencies. See the main text for more details.\label{fig:clone_impacts}}
\end{figure}

To investigate the delivery efficiencies from different positions along the impact corridor, ejecta was modeled from four locations along it. Clone A 
which had the easternmost impact longitude 40.7 East, Clone B 
which impacted closest to zero longitude (1.0 deg East), clone C 
which impacts near 60 W longitude on the Moon's leading side, and clone D 
which had the westernmost impact longitude, on the far side of the Mon at 123.7 West. The locations of the clone impacts are shown in Figure~\ref{fig:clone_impacts}. The figure shows the Lunar Reconnaissance Orbiter (LRO) Lunar Orbiter Laser Altimeter (LOLA) Global Digital Elevation Model (DEM) in a simple cylindrical projection at 118 m per pixel resolution \citep{LOLA_USGS_2014}. This dataset was imported into QGIS, an open-source geographic information system maintained by the \cite{QGIS_software}, where we applied hillshading and overlaid other basic mapping features to produce a global elevation map of the lunar surface. However the elevations are presented only for illustration purposes. The impact of clones of \yr4 are taken to occur and to produce ejecta when they intersect the lunar geoid.

\begin{figure}
\plotone{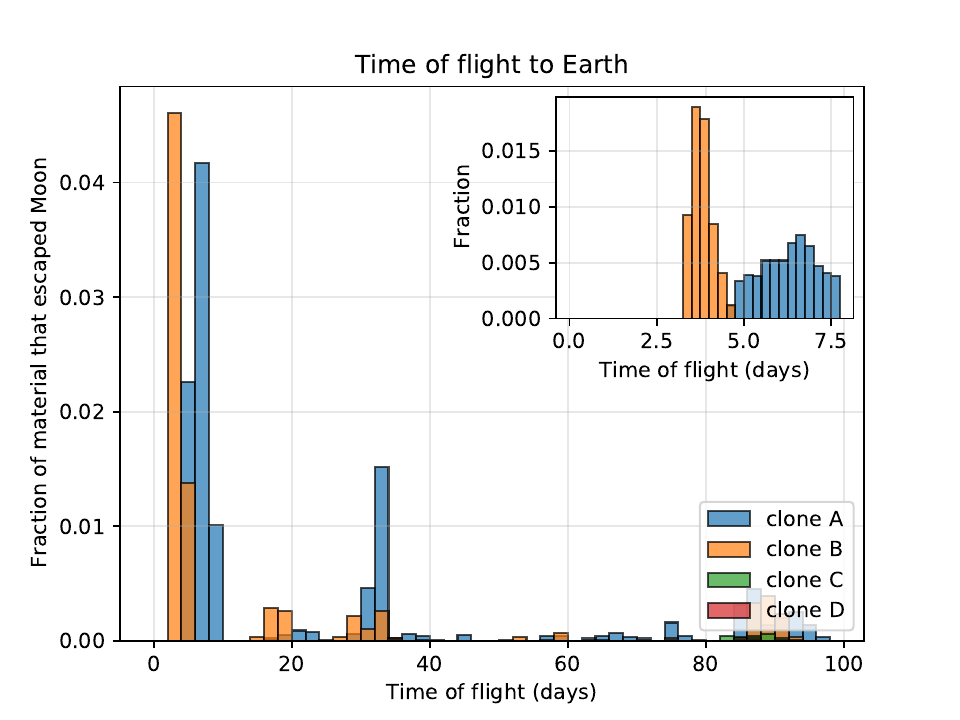}
\caption{ Time of flight from ejecta launch from the Moon until reaching the Earth for clones A through D. The inset panel shows a zoomed in view to show that first arrivals occur roughly three days after the impact for clone B and five  days for clone A. The first material from clones C and D does not arrive until over 80 days after impact.\label{fig:TOF}}
\end{figure}

Ten thousand particles were ejected from each site, and simulated forwards with a 14.4 minute time step for 100 days. The fractions of ejected material delivered to Earth were 11.9\%, 8.4\%, 0.19\% and 0.04\% respectively and the time of flight of the material is shown in Figure~\ref{fig:TOF}.  

Clones A and B produce material that arrives rapidly to Earth: in both of these cases, several percent of the total material ejected from the Moon strikes the Earth within roughly a week of its release. Material from clone B arrives more quickly, with first arrivals after only three days, while material from clone A would start arriving after five days. The material ejected from the Moon by impacts due to clones C and D only arrives at Earth after 80 days and is difficult to see in Figure~\ref{fig:TOF}: as expected, an impact on the leading side of the Moon does not deliver material quickly to Earth. Thus, the net delivery of material to Earth is quite sensitive to the impact location, with clone A or B-like impacts generating high delivery rates while C or D-like events would produce negligible amounts.

If \yr4 were to impact near the the eastern edge of the impact corridor, roughly 10\% of the $10^{7-8}$~kg of material ejected will reach the Earth quickly, the bulk arriving in just a few days. From our earlier computed particle numbers of N($>$ 10 mm) $\sim$ 10$^{5-9}$, N($>$ 1 mm) $\sim$ 10$^{11-13}$, N($>$ 100 $\mu$m) $\sim$ 10$^{14-16}$ and taking the Earth's area to be $\pi (6378~{\rm km})^2 \sim 10^{14}$m$^2$, then we can expect fluences of $10^{-10}$ to $10^{-6}$ per square meter of cm sized particles, $10^{-4}$ to $10^{-2}$ per square meter of mm-sized particles and 0.1 to 10~per square meter of $100~\mu m$ sized particles.

\section{Earth Impact Effects: Satellite and Lunar Ejecta Impact} \label{sec:effects}

Given the results of the previous section, the instantaneous meteoroid flux could -- if the asteroid impacts the Moon in a favorable spot --- reach 10 to 1000 times the background meteoroid flux at sizes which pose a hazard to astronauts and spacecraft. These sizes range from 100$\mu$m (capable of cutting exposed wires or penetrating a spacesuit \citep{cwadommcc15, Cooke_2017,Moorhead_2019_JSR}) to 1~cm (capable to mission-ending damage\footnote{\url{https://www.esa.int/Space_Safety/Space_Debris/Hypervelocity_impacts_and_protecting_spacecraft} retrieved 2025 August 5}). The standard interplanetary meteoroid flux model \citep{Moorhead_2019_JSR} predicts a mean annual flux impacting a randomly tumbling flat plate of 1 m$^2$ area near Earth of $\sim$ 1 meteoroids D $>$ 100 $\mu$m,  $\sim$ 10$^{-3}$ meteoroids D $>$ 1 mm and $\sim$ 10$^{-7}$ meteoroids D $>$ 1 cm. Our predicted flux thus represents an equivalent exposure of up to 10 years [0.1-10, 0.1-10, 0.001-10 years for each size range respectively] over just a few days.

However, the above is a mass-limited comparison - the actual damage caused by an impact depends on several factors, most notably relative speed. The mass averaged meteoroid speed near Earth is 20 km/s \citep{Grun_1985} while from Section \ref{sec:efficiency} our expected delivery velocity will be near Earth escape. Thus the relative velocity at impact on an orbiting spacecraft should vary from 4 - 18 km/s and so our mass limit may produce a factor of several less equivalent penetration damage than sporadic meteoroid impacts, but this is not significant for our order of magnitude comparison.  

This added impact exposure will result in accelerated degradation of Earth-orbiting satellites. This would occur  in Low Earth Orbit (LEO) in particular as the relative equivalent exposure of the background sporadic population will be potentially of order years to a decade. This will all occur during the few days of maximum ejecta delivery from a \yr4 impact. 

For comparison to the impact risk from orbital debris, we note that the potential particle enhancement from a \yr4 impact will greatly exceed the total impact risk from both orbital debris and background meteoroids to spacecraft in very low Earth orbit (below 270 km altitude) and at higher altitudes (above a few thousand kilometers) \citep{Moorhead_Matney_2021}. At intermediate altitudes, the relative increase in impact risk from \yr4 debris compared to orbital debris will depend on the satellite orbital inclination and impact direction. For example, near 1000 km altitude, satellites in highly inclined orbits would see an impact enhancement in the ram facing spacecraft comparable to the normal daily background orbital debris fluxes for sizes capable of penetrating 1 mm thickness of aluminum \citep{Moorhead_Matney_2021}.

For scale, it is worth noting that the current satellite cross-sectional area (active and inactive/debris) in LEO is expected to increase significantly between now and 2032\footnote{\url{https://nova.space/press-release/seven-tons-of-satellites-to-be-launched-daily-on-average-over-the-next-decade-amid-value-chain-consolidation/}}. We estimate the total cross-sectional area of all satellites in LEO and Geosynchronous Earth Orbit (GEO) in 2032 will be of order 10$^{7}$ m$^{2}$. This estimate is based on the nominal $\approx$ 100 m$^{2}$ area of the next generation of Starlink V2 mini-satellites now being launched\footnote{\url{https://spaceflightnow.com/2023/02/26/spacex-unveils-first-batch-of-larger-upgraded-starlink-satellites/}} together with the plan to orbit 30 000 by 2032 \footnote{\url{https://www.reuters.com/legal/us-court-rejects-challenges-fcc-approval-spacex-satellites-2024-07-12}/}. Here we assume that this number of Starlink satellites will be matched by roughly twice as many total satellites from all other planned megaconstellations and that these will be similar in size to the next generation of Starlink satellites, forming the bulk of the active population. While we caution this cross-sectional area estimate is at best order of magnitude, given the very large total exposed area for satellites by 2032 it becomes possible that thousands to tens of thousands of impacts from mm-sized debris ejected by a lunar impact from \yr4 will be experienced across the entire satellite fleet. Such impacts may damage satellites, but are small enough to generally not end active missions or cause breakups. 

If the orbital debris environment remains comparable in orbital distribution to today's, then we expect the relative risk to increase with increasing satellite number but that the risk will scale with altitude much as it does today.

The ejection of material from the Moon could be a serious hazard to Moon-orbiting spacecraft (e.g. Lunar Gateway), but would likely pose even greater dangers to any lunar surface operations given that most ejecta mass will accumulate across a wide swath of the moon. The precise timing and locales affected should be examined for a specific impact location should observations in 2028 result in a predicted lunar impact. 

In addition to increased flux over a few days, there is also material delivered to long-lived (months to years) Earth orbits that could pose an ongoing concern to space-based assets. These could affect operations of meteoroid-sensitive space platforms over longer time frames, though we note that the James Webb Space Telescope (JWST), which is expected to remain operational in 2032 \citep{garmatabb23} is located at the Sun-Earth $L_2$ point which is not included in this study.

We note that the production of cm-sized ejecta, which can be very damaging, is most the most uncertain as it is greatly affected by the choice of the ejecta SFD power-law exponent. Given our fluence values, the probability of impact by cm-sized lunar ejecta on any satellite surface across the entire constellation would remain of order 10\% or less. Quantifying better this upper range in the ejecta debris is important as such large impacts are generally considered to be mission-ending events and could also lead to satellite breakup in some scenarios \citep{Drolshagen_Moorhead_2019}.

Finally, the relatively low velocity lunar ejecta that will impact the atmosphere may provide an opportunity for atmospheric sampling of debris. We do not expect significant numbers of larger (decimeter-meter-sized) ejecta from the \yr4 impact given the small crater size \citep[e.g.][]{Jedicke_Alessi_Wiedner_Ghosal_Bierhaus_Granvik_2025} so meteorites are unlikely (though not impossible). However, the mass input over a week timescale from the impact of 10$^{3-4}$ Tonnes  would exceed the average daily mass input of meteoroids which is of order 10-50 T \citep{Carrillo-Sanchez_2022_mass_input} by several orders of magnitude. In this scenario almost any debris collected in the atmosphere is likely to be of lunar origin during the sedimentation timescale of dust to the Earth's surface. The resulting meteor shower could last a few days and be spectacular, though the number of visible meteors somewhat muted by the low entry speed of ejecta.

\section{Conclusions}

If \yr4 strikes the Moon in 2032, it will (statistically speaking) be the largest impact in approximately 5000~years. The delivery of ejecta escaping the Moon to near-Earth space is highly sensitive to the precise impact location, but the impact corridor as understood at this writing suggests that there is an overall 1\% chance that both 1) \yr4 will strike the Moon and 2) that the impact will have a delivery efficiency to Earth $\ge$ 10\%. The 1:100 odds of this scenario exceed the usual threshold of concern for asteroid Earth impact warning systems, which is one in a million
\citep{milchesan05,roafarche21}, indicating that this outcome is sufficiently likely to warrant further study.

If such an impact occurs, it would result in particle fluxes at Earth of 10 to 1000 times their background values, and could produce effective exposures equivalent to years in space over just a few days. The resulting meteor shower at Earth could be eye-catching, with rates orders of magnitude above usual background rates but meteor light production will be reduced by their relatively low in-atmosphere speeds. The travel time from lunar impact to Earth is typically several days but does depend on the precise location of the impact if it even occurs, which probably cannot be determined until the asteroid returns to visibility in 2028. Material persisting in Earth orbit for longer times could also present a hazard. 

Our analysis highlights that issues of planetary defense extend beyond just the effects of impacts on Earth's surface. 
The NASA Planetary Defense Strategy and Action Plan (2023)\footnote{\url{https://www.nasa.gov/wp-content/uploads/2023/06/nasa_-_planetary_defense_strategy_-_final-508.pdf?emrc=37bb97}} states on page 11 that 
"Planetary defense encompasses all the capabilities needed to detect and warn of potential asteroid or comet impacts {\it with Earth} and then to either prevent or mitigate their possible effects." (emphasis added).  The Moon itself is not explicitly mentioned in that 42 page document except to note that space situational awareness (SSA) systems acting in support of missions to the Moon could also be used for Near-Earth Object (NEO) detection and tracking. Though this is admittedly only one of many documents addressing the topic of planetary defense, current thinking has as a rule considered it to pertain only to the mitigation of asteroid impacts with the Earth itself.

At the same time, the Moon does already effectively figure in basic planetary defense considerations, as (for example) impact probabilities with the Moon for NEOs like \yr4 are routinely calculated and published. However, the parameters for lunar defense are different for the Earth. The size range of interest is also different for the Moon because of its airless surface. Objects to small to survive entry into Earth's atmosphere could strike the Moon's surface at hypervelocity, dispersing material at high speed over a wide area and even ejecting material into space. And should historically important sites be part of the equation? \yr4 is unlikely to strike near any landed lunar missions, with the possible exception of Surveyor 7 \citep{sto25} but what if the impact corridor passed near the Apollo 11 site?   These are all issues which require additional thought. We suggest that the global space community consider extending the formal definition and scope of planetary defense beyond the confines of planet Earth itself to the wider domain of near-Earth and lunar space, where hardware assets and spacefaring humans could suffer adverse affects from NEO impacts under situations not addressed by traditional Earth impact scenarios.

\begin{acknowledgments}
This study was supported in part by the NASA Meteoroid Environment Office under Cooperative Agreement No. 80NSSC24M0060 and by the Natural Sciences and Engineering Research Council of Canada (NSERC) Discovery Grant program (grants No. RGPIN-2018-05659 and RGPIN-2024-05200)
\end{acknowledgments}

\software{SpiceyPy \citep{Annex2020}, SPICE \citep{Acton2018}, DE440 \citep{parfolwil21}, QGIS \citep{QGIS_software}}



\bibliography{2024YR4}{}

\begin{thebibliography}{}
\expandafter\ifx\csname natexlab\endcsname\relax\def\natexlab#1{#1}\fi
\providecommand{\url}[1]{\href{#1}{#1}}
\providecommand{\dodoi}[1]{doi:~\href{http://doi.org/#1}{\nolinkurl{#1}}}
\providecommand{\doeprint}[1]{\href{http://ascl.net/#1}{\nolinkurl{http://ascl.net/#1}}}
\providecommand{\doarXiv}[1]{\href{https://arxiv.org/abs/#1}{\nolinkurl{https://arxiv.org/abs/#1}}}

\bibitem[{C.~H. Acton {et~al.}(2018)Acton, Bachman, Semenov, \&
  Wright}]{Acton2018}
Acton, C.~H., Bachman, N.~J., Semenov, B.~G., \& Wright, E.~D. 2018,
  \bibinfo{title}{A look toward the future in the handling of space science
  mission geometry,} Planetary and Space Science, 150, 9,
  \dodoi{10.1016/j.pss.2017.02.013}

\bibitem[{A.~M. Annex(2020)Annex}]{Annex2020}
Annex, A.~M. 2020, \bibinfo{title}{SpiceyPy: A Pythonic Wrapper for the SPICE
  Toolkit,} Journal of Open Source Software, 5, 2050,
  \dodoi{10.21105/joss.02050}

\bibitem[{G.~D. Bart \& H.~J. Melosh(2010)Bart \& Melosh}]{Bart_Melosh_2010b}
Bart, G.~D., \& Melosh, H.~J. 2010, \bibinfo{title}{Distributions of boulders
  ejected from lunar craters,} Icarus, 209, 337–357,
  \dodoi{10.1016/j.icarus.2010.05.023}

\bibitem[{J.~D. Carrillo-Sanchez {et~al.}(2022)Carrillo-Sanchez, Janches,
  Plane, Pokorný, Sarantos, Crismani, Feng, \&
  Marsh}]{Carrillo-Sanchez_2022_mass_input}
Carrillo-Sanchez, J.~D., Janches, D., Plane, J., {et~al.} 2022,
  \bibinfo{title}{A Modeling Study of the Seasonal, Latitudinal, and Temporal
  Distribution of the Meteoroid Mass Input at Mars: Constraining the Deposition
  of Meteoric Ablated Metals in the Upper Atmosphere,} Planetary Science
  Journal, 3, 239, \dodoi{10.3847/PSJ/ac8540}

\bibitem[{J.~D. {Castro-Cisneros} {et~al.}(2025){Castro-Cisneros}, {Malhotra},
  \& {Rosengren}}]{casmalros25}
{Castro-Cisneros}, J.~D., {Malhotra}, R., \& {Rosengren}, A.~J. 2025,
  \bibinfo{title}{{Lunar impact ejecta flux on the Earth},} \icarus, 438,
  116606, \dodoi{10.1016/j.icarus.2025.116606}

\bibitem[{W. Cooke {et~al.}(2017)Cooke, Matney, Moorhead, \&
  Vavrin}]{Cooke_2017}
Cooke, W., Matney, M., Moorhead, A.~V., \& Vavrin, A. 2017, \bibinfo{title}{A
  comparison of damaging meteoroid and orbital debris fluxes in Earth orbit,}
  in European Conference on Space Debris No. M17-5915

\bibitem[{C.~D. Cwalina {et~al.}(2015)Cwalina, Dombrowski, McCutcheon,
  Christiansen, \& Wagner}]{cwadommcc15}
Cwalina, C.~D., Dombrowski, R.~D., McCutcheon, C.~J., Christiansen, E.~L., \&
  Wagner, N.~J. 2015, \bibinfo{title}{MMOD Puncture Resistance of EVA Suits
  with Shear Thickening Fluid (STF) – Armortm Absorber Layers,} Procedia
  Engineering, 103, 97, \dodoi{https://doi.org/10.1016/j.proeng.2015.04.014}

\bibitem[{G. Drolshagen \& A. Moorhead(2019)Drolshagen \&
  Moorhead}]{Drolshagen_Moorhead_2019}
Drolshagen, G., \& Moorhead, A. 2019, The Meteoroid Impact Hazard for
  Spacecraft, ed. D.~J. Asher, G.~O. Ryabova, \& M.~D. Campbell-Brown,
  Cambridge Planetary Science (Cambridge: Cambridge University Press),
  255–274, \dodoi{10.1017/9781108606462.019}

\bibitem[{E. {Everhart}(1985){Everhart}}]{eve85}
{Everhart}, E. 1985, \bibinfo{title}{{An efficient integrator that uses
  Gauss-Radau spacings},} in Dynamics of Comets: Their Origin and Evolution,
  ed. A.~{Carusi} \& G.~B. {Valsecchi} (Dordrecht: Kluwer), 185--202

\bibitem[{J.~P. {Gardner} {et~al.}(2023){Gardner}, {Mather}, {Abbott}, {Abell},
  {Abernathy}, {Abney}, {Abraham}, {Abraham}, {Abul-Huda}, {Acton}, {Adams},
  {Adams}, {Adler}, {Adriaensen}, {Aguilar}, {Ahmed}, {Ahmed}, {Ahmed},
  {Albat}, {Albert}, {Alberts}, {Aldridge}, {Allen}, {Allen}, {Altenburg},
  {Altunc}, {Alvarez}, {{\'A}lvarez-M{\'a}rquez}, {Alves de Oliveira},
  {Ambrose}, {Anandakrishnan}, {Andersen}, {Anderson}, {Anderson}, {Anderson},
  {Anderson}, {Aprea}, {Archer}, {Arenberg}, {Argyriou}, {Arribas}, {Artigau},
  {Arvai}, {Atcheson}, {Atkinson}, {Averbukh}, {Aymergen}, {Bacinski},
  {Baggett}, {Bagnasco}, {Baker}, {Balzano}, {Banks}, {Baran}, {Barker},
  {Barrett}, {Barringer}, {Barto}, {Bast}, {Baudoz}, {Baum}, {Beatty},
  {Beaulieu}, {Bechtold}, {Beck}, {Beddard}, {Beichman}, {Bellagama}, {Bely},
  {Berger}, {Bergeron}, {Bernier}, {Bertch}, {Beskow}, {Betz}, {Biagetti},
  {Birkmann}, {Bjorklund}, {Blackwood}, {Blazek}, {Blossfeld}, {Bluth},
  {Boccaletti}, {Boegner}, {Bohlin}, {Boia}, {B{\"o}ker}, {Bonaventura},
  {Bond}, {Bosley}, {Boucarut}, {Bouchet}, {Bouwman}, {Bower}, {Bowers},
  {Bowers}, {Boyce}, {Boyer}, {Boyer}, {Boyer}, {Boyer}, {Bradley}, {Brady},
  {Brandl}, {Brannen}, {Breda}, {Bremmer}, {Brennan}, {Bresnahan}, {Bright},
  {Broiles}, {Bromenschenkel}, {Brooks}, {Brooks}, {Brown}, {Brown}, {Brown},
  {Bruce}, {Bryson}, {Bujanda}, {Bullock}, {Bunker}, {Bureo}, {Burt}, {Bush},
  {Bushouse}, {Bussman}, {Cabaud}, {Cale}, {Calhoon}, {Calvani}, {Canipe},
  {Caputo}, {Cara}, {Carey}, {Case}, {Cesari}, {Cetorelli}, {Chance},
  {Chandler}, {Chaney}, {Chapman}, {Charlot}, {Chayer}, {Cheezum}, {Chen},
  {Chen}, {Cherinka}, {Chichester}, {Chilton}, {Chittiraibalan}, {Clampin},
  {Clark}, {Clark}, {Clark}, {Claybrooks}, {Cleveland}, {Cohen}, {Cohen},
  {Col{\'o}n}, {Coleman}, {Colina}, {Comber}, {Comeau}, {Comer}, {Conde Reis},
  {Connolly}, {Conroy}, {Contos}, {Contreras}, {Cook}, {Cooper}, {Cooper},
  {Correia}, {Correnti}, {Cossou}, {Costanza}, {Coulais}, {Cox}, {Coyle},
  {Cracraft}, {Crew}, {Curtis}, {Cusveller}, {Da Costa Maciel}, {Dailey},
  {Daugeron}, {Davidson}, {Davies}, {Davis}, {Davis}, {Day}, {de Chambure}, {de
  Jong}, {De Marchi}, {Dean}, {Decker}, {Delisa}, {Dell}, \&
  {Dellagatta}}]{garmatabb23}
{Gardner}, J.~P., {Mather}, J.~C., {Abbott}, R., {et~al.} 2023,
  \bibinfo{title}{{The James Webb Space Telescope Mission},} \pasp, 135,
  068001, \dodoi{10.1088/1538-3873/acd1b5}

\bibitem[{B.~J. {Gladman} {et~al.}(1995){Gladman}, {Burns}, {Duncan}, \&
  {Levison}}]{glaburdun95}
{Gladman}, B.~J., {Burns}, J.~A., {Duncan}, M.~J., \& {Levison}, H.~F. 1995,
  \bibinfo{title}{{The dynamical evolution of lunar impact ejecta.},} \icarus,
  118, 302, \dodoi{10.1006/icar.1995.1193}

\bibitem[{E. Grün {et~al.}(1985)Grün, Zook, Fechtig, \& Giese}]{Grun_1985}
Grün, E., Zook, H., Fechtig, H., \& Giese, R. 1985,
  \bibinfo{title}{Collisional balance of the meteoritic complex,} Icarus, 62,
  244–272

\bibitem[{K.~A. Holsapple(1993)Holsapple}]{Holsapple_1993}
Holsapple, K.~A. 1993, \bibinfo{title}{The Scaling of Impact Processes in
  Planetary Sciences,} Annual Review of Earth and Planetary Sciences, 21,
  333–373, \dodoi{10.1146/annurev.ea.21.050193.002001}

\bibitem[{K.~A. Holsapple \& R.~M. Schmidt(1982)Holsapple \&
  Schmidt}]{Holsapple_Schmidt_1982}
Holsapple, K.~A., \& Schmidt, R.~M. 1982, \bibinfo{title}{On the scaling of
  crater dimensions: 2. Impact processes,} Journal of Geophysical Research:
  Solid Earth, 87, 1849–1870, \dodoi{10.1029/JB087iB03p01849}

\bibitem[{K.~A. Holsapple \& R.~M. Schmidt(1987)Holsapple \&
  Schmidt}]{Holsapple_Schmidt_1987}
Holsapple, K.~A., \& Schmidt, R.~M. 1987, \bibinfo{title}{Point source
  solutions and coupling parameters in cratering mechanics,} Journal of
  Geophysical Research: Solid Earth, 92, 6350–6376,
  \dodoi{10.1029/JB092iB07p06350}

\bibitem[{K.~R. Housen \& K.~A. Holsapple(2011)Housen \&
  Holsapple}]{Housen_Holsapple_2011}
Housen, K.~R., \& Holsapple, K.~A. 2011, \bibinfo{title}{Ejecta from impact
  craters,} Icarus, 211, 856–875, \dodoi{10.1016/j.icarus.2010.09.017}

\bibitem[{K.~R. Housen {et~al.}(1983)Housen, Schmidt, \&
  Holsapple}]{Housen_Schmidt_Holsapple_1983}
Housen, K.~R., Schmidt, R.~M., \& Holsapple, K.~A. 1983, \bibinfo{title}{Crater
  ejecta scaling laws: Fundamental forms based on dimensional analysis,}
  Journal of Geophysical Research: Solid Earth, 88, 2485–2499,
  \dodoi{10.1029/JB088iB03p02485}

\bibitem[{R. Jedicke {et~al.}(2025)Jedicke, Alessi, Wiedner, Ghosal, Bierhaus,
  \& Granvik}]{Jedicke_Alessi_Wiedner_Ghosal_Bierhaus_Granvik_2025}
Jedicke, R., Alessi, E.~M., Wiedner, N., {et~al.} 2025, \bibinfo{title}{The
  steady state population of Earth’s minimoons of lunar provenance,} Icarus,
  438, 116587, \dodoi{10.1016/j.icarus.2025.116587}

\bibitem[{R. {Jedicke} {et~al.}(2025){Jedicke}, {Alessi}, {Wiedner}, {Ghosal},
  {Bierhaus}, \& {Granvik}}]{jedalewie25J}
{Jedicke}, R., {Alessi}, E.~M., {Wiedner}, N., {et~al.} 2025,
  \bibinfo{title}{{The steady state population of Earth's minimoons of lunar
  provenance},} \icarus, 438, 116587, \dodoi{10.1016/j.icarus.2025.116587}

\bibitem[{Y. {Jiao} {et~al.}(2024){Jiao}, {Cheng}, {Huang}, {Asphaug},
  {Gladman}, {Malhotra}, {Michel}, {Yu}, \& {Baoyin}}]{jiachehua24}
{Jiao}, Y., {Cheng}, B., {Huang}, Y., {et~al.} 2024, \bibinfo{title}{{Asteroid
  Kamo`oalewa's journey from the lunar Giordano Bruno crater to Earth 1:1
  resonance},} Nature Astronomy, 8, 819, \dodoi{10.1038/s41550-024-02258-z}

\bibitem[{M.~A. {Kreslavsky} \& E. {Asphaug}(2014){Kreslavsky} \&
  {Asphaug}}]{kreasp14}
{Kreslavsky}, M.~A., \& {Asphaug}, E. 2014, \bibinfo{title}{{Direct Delivery of
  Lunar Impact Ejecta to the Earth},} in 45th Annual Lunar and Planetary
  Science Conference, Lunar and Planetary Science Conference, 2455

\bibitem[{H.~J. Melosh(1985)Melosh}]{Melosh_1985}
Melosh, H.~J. 1985, \bibinfo{title}{Ejection of rock fragments from planetary
  bodies,} Geology, 13, 144–148,
  \dodoi{10.1130/0091-7613(1985)13<144:EORFFP>2.0.CO;2}

\bibitem[{H.~J. {Melosh}(1989){Melosh}}]{mel89}
{Melosh}, H.~J. 1989, {Impact cratering : a geologic process}

\bibitem[{A. {Milani} {et~al.}(2005){Milani}, {Chesley}, {Sansaturio},
  {Tommei}, \& {Valsecchi}}]{milchesan05}
{Milani}, A., {Chesley}, S.~R., {Sansaturio}, M.~E., {Tommei}, G., \&
  {Valsecchi}, G.~B. 2005, \bibinfo{title}{{Nonlinear impact monitoring: line
  of variation searches for impactors},} \icarus, 173, 362,
  \dodoi{10.1016/j.icarus.2004.09.002}

\bibitem[{A. Moorhead {et~al.}(2019)Moorhead, Egal, Brown, Moser, \&
  Cooke}]{Moorhead_2019_JSR}
Moorhead, A., Egal, A., Brown, P.~G., Moser, D.~E., \& Cooke, W. 2019,
  \bibinfo{title}{Meteor Shower Forecasting in Near-Earth Space,} Journal of
  Spacecraft and Rockets, 56, 1531–1545, \dodoi{10.2514/1.A34416}

\bibitem[{A. Moorhead \& M. Matney(2021)Moorhead \&
  Matney}]{Moorhead_Matney_2021}
Moorhead, A., \& Matney, M. 2021, \bibinfo{title}{The ratio of hazardous
  meteoroids to orbital debris in near-Earth space,} Advances in Space
  Research, 67, 384–392, \dodoi{https://doi.org/10.1016/j.asr.2020.09.015}

\bibitem[{G. Neukum {et~al.}(2001)Neukum, Ivanov, \&
  Hartmann}]{Neukum_Ivanov_Hartmann_2001}
Neukum, G., Ivanov, B., \& Hartmann, W.~K. 2001, \bibinfo{title}{Cratering
  records in the inner solar system in relation to the lunar reference system,}
  Space Science Reviews, 96, 55–86, \dodoi{10.1023/A:1011989004263}

\bibitem[{R.~S. {Park} {et~al.}(2021){Park}, {Folkner}, {Williams}, \&
  {Boggs}}]{parfolwil21}
{Park}, R.~S., {Folkner}, W.~M., {Williams}, J.~G., \& {Boggs}, D.~H. 2021,
  \bibinfo{title}{{The JPL Planetary and Lunar Ephemerides DE440 and DE441},}
  \aj, 161, 105, \dodoi{10.3847/1538-3881/abd414}

\bibitem[{ {QGIS Development Team}(2025){QGIS Development
  Team}}]{QGIS_software}
{QGIS Development Team}. 2025, QGIS Geographic Information System, QGIS
  Association.
\newblock \url{https://www.qgis.org}

\bibitem[{A.~S. Rivkin {et~al.}(2025)Rivkin, Mueller, MacLennan, Holler,
  Burdanov, de~Wit, Pravec, Micheli, Devogele, Conversi, Thomas, Farnocchia,
  Dotson, Wheeler, Hammel, Milam, de~Leon, \&
  Glantzberg}]{Rivkin_2025_JWST_2025}
Rivkin, A.~S., Mueller, T., MacLennan, E., {et~al.} 2025, \bibinfo{title}{JWST
  Observations of Potentially Hazardous Asteroid 2024 YR4,} Research Notes of
  the AAS, 9, 70, \dodoi{10.3847/2515-5172/adc6f0}

\bibitem[{J. {Roa} {et~al.}(2021){Roa}, {Farnocchia}, \&
  {Chesley}}]{roafarche21}
{Roa}, J., {Farnocchia}, D., \& {Chesley}, S.~R. 2021, \bibinfo{title}{{A Novel
  Approach to Asteroid Impact Monitoring},} \aj, 162, 277,
  \dodoi{10.3847/1538-3881/ac193f}

\bibitem[{K.~N. Singer {et~al.}(2020)Singer, Jolliff, \&
  McKinnon}]{Singer_Jolliff_McKinnon_2020}
Singer, K.~N., Jolliff, B.~L., \& McKinnon, W.~B. 2020, \bibinfo{title}{Lunar
  Secondary Craters and Estimated Ejecta Block Sizes Reveal a Scale-Dependent
  Fragmentation Trend,} Journal of Geophysical Research: Planets, 125,
  e2019JE006313, \dodoi{10.1029/2019JE006313}

\bibitem[{E.~J. {Speyerer} {et~al.}(2011){Speyerer}, {Robinson}, {Denevi}, \&
  {LROC Science Team}}]{sperobden11}
{Speyerer}, E.~J., {Robinson}, M.~S., {Denevi}, B.~W., \& {LROC Science Team}.
  2011, \bibinfo{title}{{Lunar Reconnaissance Orbiter Camera Global
  Morphological Map of the Moon},} in 42nd Annual Lunar and Planetary Science
  Conference, Lunar and Planetary Science Conference, 2387

\bibitem[{P. Stooke(2025)Stooke}]{sto25}
Stooke, P. 2025, private communication,,
  \url{https://publish.uwo.ca/~pjstooke/moon-sites-map.jpg}

\bibitem[{ {USGS Astrogeology Science Center}(2014){USGS Astrogeology Science
  Center}}]{LOLA_USGS_2014}
{USGS Astrogeology Science Center}. 2014, {LRO LOLA 118m Global Lunar DEM},
  \url{https://astrogeology.usgs.gov/search/map/moon_lro_lola_dem_118m}

\bibitem[{A.~M. {Vickery}(1987){Vickery}}]{vic87}
{Vickery}, A.~M. 1987, \bibinfo{title}{{Variation in ejecta size with ejection
  velocity},} \grl, 14, 726, \dodoi{10.1029/GL014i007p00726}

\bibitem[{R.~N. Watkins {et~al.}(2019)Watkins, Jolliff, Mistick, Fogerty,
  Lawrence, Singer, \& Ghent}]{watkins_boulder_2019}
Watkins, R.~N., Jolliff, B.~L., Mistick, K., {et~al.} 2019,
  \bibinfo{title}{Boulder {Distributions} {Around} {Young}, {Small} {Lunar}
  {Impact} {Craters} and {Implications} for {Regolith} {Production} {Rates} and
  {Landing} {Site} {Safety},} Journal of Geophysical Research: Planets, 124,
  2754, \dodoi{10.1029/2019JE005963}

\end{thebibliography}
\bibliographystyle{aasjournalv7}



\end{document}